# Fast spin dynamics in hexagonal arrays of Fe atoms on metallic surface

Tamene R. Dasa[*,1], Valeri S. Stepanyuk[†,2]

[1]Department of Materials Science and Engineering, The University of Tennessee Knoxville, USA

[2]Max-Planck-Institut für Mikrostrukturphysik, Weinberg 2, D-06120 Halle, Germany

Email:[*]*tdasa@utk.edu,*[†]*valeri.stepanyuk@mpi-halle.mpg.de*

## Abstract

Understanding the nature of magnetic interactions in ultra-small magnetic ensembles and their intrinsic properties is vital to uncover the dynamics therein. In this study, we reveal the spin dynamics of hexagonally arranged Fe atoms on metallic surface that are triggered by magnetic pulse. The switching process among various spin configurations and their relative magnetic order is tuned by the amplitude and duration of the magnetic pulse. Even more we observe a parity effect in the switching time as the size of the cluster varies in which even number of Fe atoms shows faster dynamics. The changes in the multistable magnetic states and switching times are explained by using the relaxation of the exchange and anisotropy energies in time.

Comprehensive understanding of magnetic clusters at the atomic scale is useful in an effort to identify the magnetic entities that can be incorporated into faster and smaller spintronic devices. These components are expected to be stable against thermal fluctuations. For instance a relatively stable atomic superlattices or arrays are investigated in experimental studies.[1] On the other hand, in order to realize functional magnetic devices, the spin signal of magnetic units should be altered with external stimuli or be propagated in magnetic medium.[2] Under such circumstances the finding out the nature of the magnetization dynamics is vital. This leads us yet to another crucial spin phenomena; the temporal evolution of magnetization. In fact, similar magnetization dynamics does exist during experimental measurements.[3,4]

Controlling spin dynamics could allow us to determine transition among magnetic states, in adsorbed clusters, which is sensitive to a magnetic environment as well as the magnetic anisotropy energy. Moreover, the relative alignments of magnetic moments within the cluster strongly affects their dynamical properties.[4] For instance antiferromagnetically aligned materials are relatively insensitive to an external magnetic fields and have relatively faster spin dynamics as compared to the ferromagnetic counterparts.[5] In fact, antiffermagnetic chains could be used as interconnects in spin logic devices[1] as well as memory units[3,4] that can be mentioned as few achievements in antiferromagnetic spintronics. In both applications the spin dynamics plays significant role for setting the final magnetic state and the rate of information transmission.[1]

A fascinating way of engineering antiferromagnetic arrays of Fe atoms on metallic surfaces is realized employing scanning tunneling spectroscopy technique.[6] Such configuration of magnetic atoms on metal surfaces is stabilized by surface states[7,8], or in some cases combined with organic structures.[9,10] Understanding the spin trajectories of atomic arrays on surfaces under external magnetic field is vital for the aforementioned properties of antiferromagnets. In particular, the spin dynamics of atomic scale magnets is almost unexplored where most studies are restricted to bulk materials. Multiple studies have employed a short laser pulse as external stimuli to study the magnetization dynamics of thin film antiferromagnets,[13,15,16] in which the magnetic component that derives the excitation process and inertial switching behavior is observed.[13,14,17] Even more, constant magnetic field is used for controlling magnetic order in nanostructures and other efficient means is being explored.[4,11,12]

In this study we reveal reasonably fast spin dynamics in hexagonal arrays of Fe atoms on Cu(111) surface by using short magnetic pulses. The switching scenarios among various magnetic states as a function of the amplitude and duration of the magnetic pulse are examined. The trajectories of the spin and all components of the total energy are analyzed in order to explain the switching processes. Moreover, we observe a parity effect in the switching time for different sizes of antiferromagnetic clusters that are exposed to a moderate magnetic pulse.

The spin dynamics of 2D magnetic clusters calculated using semi-classical technique by solving the atomistic Landau-Lifshitz-Gilbert (LLG) Equation,[18] written as,

$$\frac{\partial S_i}{\partial t} = \frac{\gamma}{(1+\lambda^2)} S_i \times [H_{i,eff} + \lambda(S_i \times H_{i,eff})]$$

Basically, LLG equation evaluates the atomic resolved magnetic trajectories within an on-site effective magnetic field. The total Hamiltonian (H) is given as $H=H_{exc}+H_{MAE}+H_{app}$. The Zeeman term is modified in such a way that it incorporates the constant magnetic field and time dependent magnetic field in which the latter mimics the magnetic pulse. The absolute value of the gyromagnetic ratio is $\gamma = 1.76 \times 10^{11}$ $T^{-1}s^{-1}$. The local spin moments are denoted with unit vector $S_i$ and $S_j$ obtained from the realistic atomic moments as $S_i = m_s / |m_s|$. We have considered a damping parameter of $\lambda = 0.006$. A small temperature of 0.2 K is added in order to include thermal effects as well as to introduce non-zero processional term at the beginning of the spin dynamics.

In an effort to reveal the responses of the magnetic states of atomic clusters to magnetic stimuli, here we present the evolution of atomic resolved magnetization of hexagonal arrays of Fe atoms on Cu(111) surfaces for various magnitude of a magnetic pulse. Our study is motivated by experimental work on atom by atom assembly of different arrays of Fe atoms in various structures and the manipulation of their metastable sates with external magnetic field.[6] In particular, the interest lies on the flower like magnetic structure that have four fold degenerate ground states without magnetic field. Evidently, the degeneracy of the magnetic state in such clusters can be lifted in the presence of magnetic field.[6] Moreover, we would like to show that one of the magnetic states can be favored over the other using short magnetic pulse.

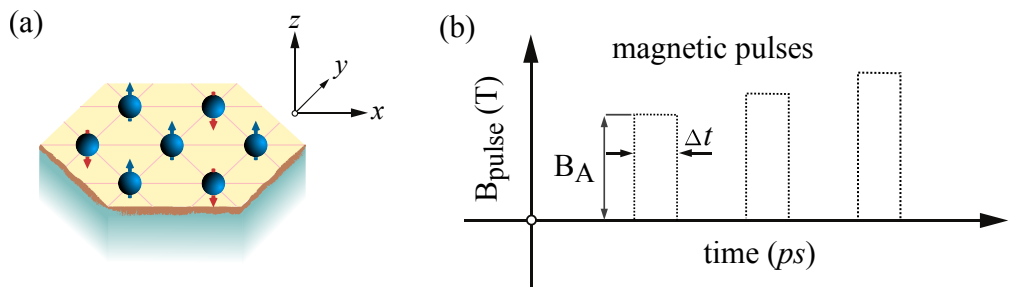

Fig. 1. (a) The schematic representation of the hexagonal array of Fe atoms on Cu(111) surface. (b) The direction, along x-axis, amplitude ($B_A$) and duration of the magnetic pulse ($\Delta t$) is also depict. The easy axis of magnetization is directed to out-of-plane direction and at zero magnetic field the pairwise exchange coupling is dominantly antiferromagnetic.

Actually, a very well-focused electron beam has been experimentally used to produce short magnetic field pulses of several Tesla for about 2 to 5 ps and magnetization reversal in nanostructures can be induced therewith.[17] The Fe atoms on Cu(111) are interspaced by 9.2 Å and a local magnetic moment of 3.2 $\mu_B$ is used for Fe atom, as shown in Fig. 1 (a). In the spin dynamics simulation a pairwise exchange coupling of 0.1 meV is used among the atomic sites, which is adopted from the experimental results.[6] The values of MAE (1 meV per Fe atom) obtained from our electronic structure calculations are in agreement with the experimental results.[6] These calculations are performed on the basis of density functional theory within the projector augmented wave technique.[19] The local spin density approximation[20] and plane wave basis sets are used for the exchange correlation interactions and the Kohn-Sham wave function, respectively.

The validity of employing such semi-classical approach for studying the spin trajectories of atomic clusters is discussed as follows. The quantum spin tunneling, which is one flavor of quantum spin dynamics, seems not to exist in single ad-atoms where the exchange interaction with itinerant electrons is enhanced[22]. Thus the superpositions between the spin states is minimized. Furthermore, in such highly-spaced Fe atoms where the transverse anisotropy is appreciably reduced, it is shown that the QST phenomena does not exit.[23] Secondly, considering the thermal effects on the spin dynamics, according to Gauyacq and Lorente,[25] lead us to the conclusion that hexagonally arranged Fe atoms on metallic sur- face will certainly lie in the classical regime, leading to spontaneous magnetization. The localized spin dynamics equation that are employed in this study assumes the Born-Oppenheimer approximation implying that the motions of the atoms is much slower than the fast-moving electrons.[26] Hence, employing such techniques for magnetization process that scales down to few picoseconds is valid.

In this work, a rectangular magnetic pulses are used that are applied for time scale of much shorter than the spin relaxation time. In Fig. 2, we present the saturation magnetization of selected atoms with respect to the amplitude of the magnetic pulse. The amplitude of the magnetic pulse increases linearly and for each amplitude of the magnetic pulse has a duration of $\Delta t = 5$ ps and spin dynamics is performed at each magnetic field, see Fig.1 (b). At zero magnetic field we start from one of the degenerate AF configurations and the spin dynamics simulation continues until we reach to new metastable state or saturation magnetization. In most cases the magnetization of the cluster relaxes to the z-axis, either to spin up or spin down, accompanied by the precession of the x and y spin components. Thereafter a new simulation will begin from the latter state by increasing the magnitude of the pulse until we reach to the next spin state.

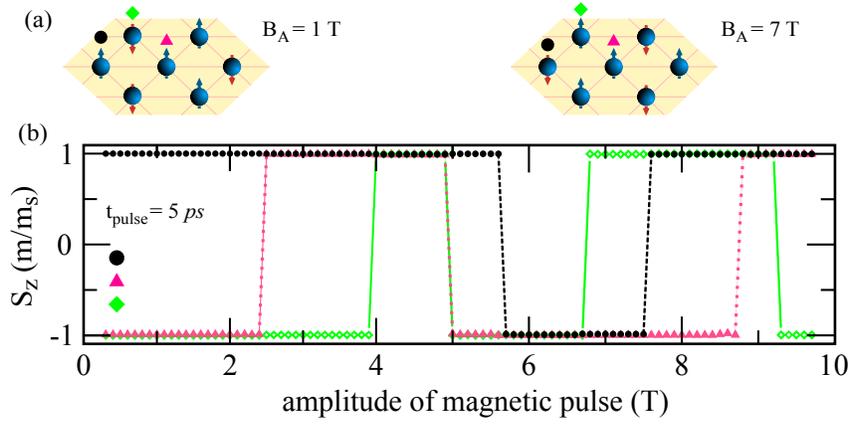

Fig. 2. (a) Multiple magnetic states of the hexagonal array of Fe atoms are explored, as well as the transition among the states, by using magnetic pulse, $B_A$. (b) The saturated magnetization of selected atoms in hexagonal array of Fe atoms as function of the amplitude of the magnetic pulse, applied for finite time of 5 ps. The z-component magnetization is presented for two non-similar Fe atoms and central Fe atom assigned with circle, diamond and triangle, respectively.

For relatively small amplitude of the magnetic pulse of $B_A$ = 2.4 T the central atom switches from spin up configuration to spin down, driving the system to another metastable state. The cluster remains AF order for most magnetic fields, except within the range of 5 T to 6.8 T it changes to ferromagnetic (FM) order. Then at 6.8 T all magnetic moments reverse their direction, relative to the initial spin directions, suggesting the possibility of tuning the Neel states in such hexagonal arrays. Analyzing magnetic states for all magnetic fields, one can see that the Fe atoms in the array undergo multiple switching among its spin configuration. In experimental study similar way of inducing non-degenerate magnetic structures by external magnetic field have been observed, which agrees with our calculations. Hence, we showed that by varying only the amplitude of magnetic pulse the magnetic configuration of 2D nanostructures can be sequentially traced. The spin dynamics does also depend on the duration of the magnetic pulse. For the same amplitude when the cluster is exposed to even shorter time of 1.2 ps the final magnetic configuration is dominated by anti-ferromagnetic pair-wise interaction. Moreover, it is followed by magnetization reversal and the spins relax in much slower mode, leading to longer switching time of 40 ps.

Such tuning of magnetic states can be inferred by closely examining the spin and energy relaxations in time. In Fig. 3 the spin trajectories of representative atomic spins from their initial to final spin states are depicted. The amplitude of the magnetic field is 5 T which is applied for 5 ps and initially the Fe atoms have ferromagnetic coupling. Once the system is exposed to short magnetic pulse the relative magnetic order saturates very fast to antiferromagnetic alignment. In this case the switching time is found to be 5 ps and the pairwise exchange coupling ends up to be antiferromagnet but the central Fe atom takes relatively longer time to reach to a complete saturation. The color code in Fig. 3 (a) represents the z-component of the magnetization of each atom in the hexagonal lattice. The first two snap shots are before the magnetic pulse is switched off, specifically the pic- tures are taken just after the spin dynamics begins and at 3.5 ps. The latter one depicts a non-collinear magnetic configuration and the plot at 10 ps is an exemplary of a saturated magnetic order.

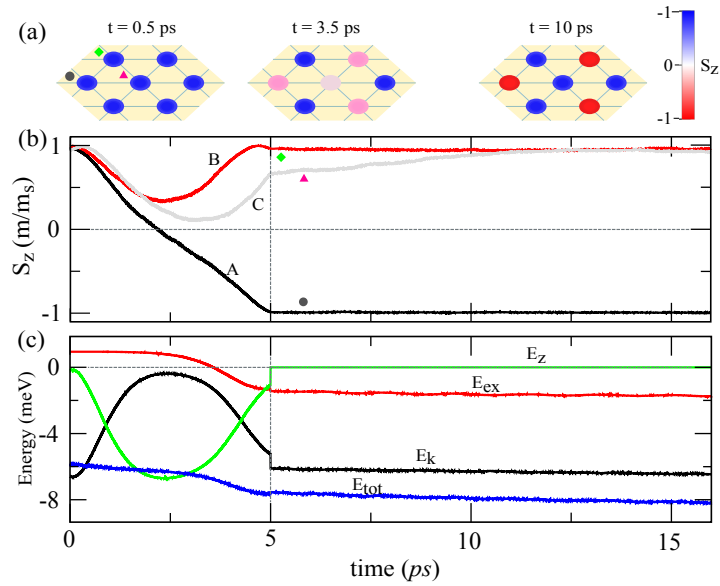

Fig. 3. The spin trajectories of Fe atoms and the evolution of energies in time for hexagonally arranged Fe atoms on Cu(111) surface. (a) The atomic resolved magnetic moments (z-components) at three steps of the spin trajectories represented with color code. (b) The dynamical switching in the z-component of the magnetic moments of Fe atoms in time. A magnetic pulse of 5 T is applied close to the y-axis ($\theta = 85$, $\varphi = 87$) for 5 ps. (c) The total energy and its components as function of time. The Zeeman energy ($E_Z$), exchange ($E_{ex}$) and anisotropy ($E_k$) are denoted, respectively.

There is a considerable dynamics of magnetic energy transfer from the short magnetic pulse to the exchange energy and the anisotropy energy. The phenomena that derives such magnetization reversal is related to the inertia effect during the magnetic field,[13,14] and can also be explained from energy trajectories. In order to observe the latter phenomena and hence explain the spin dynamics, in Fig.3 (c) we plot the changes of magnetic energy in time. Analogous to the spin dynamics the trajectories of the magnetic energy strongly vary in time until the magnetic pulse is off and saturates shortly afterward. The spin trajectories mainly depend on the energy redistribution just before the end of magnetic pulse. In this time range a substantial energy is transferred from the Zeeman energy to the anisotropy and exchange energies. The energy transferred to the exchange is strong enough to induce magnetic reversal in some of the Fe atoms and suppress further damping in the post magnetic pulse regime. Certainly, varying either the amplitude or duration of the magnetic pulse one can drastically change the spin dynamics end the final magnetic configuration.

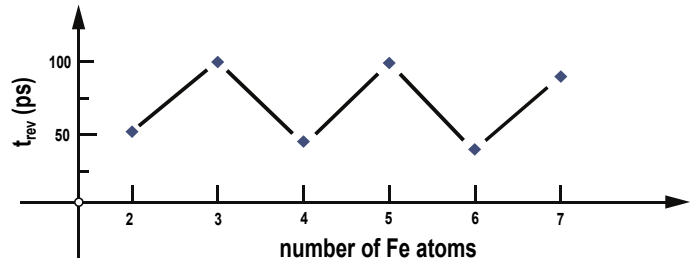

Fig. 4. The interplay between the switching time and the size of Fe clusters on Cu(111) surface. In all cases, atomic spins start from same magnetic configuration and a magnetic pulse of 6.8 T is applied parallel to the plane. It clearly shows the presence of parity effect on the spin dynamics where slower switching time is observed for odd number of Fe atoms.

It is important to note that the spin trajectories de- pend on the number of pair-wise exchange coupling within a cluster and the initial magnetic state where the magnetic dynamics started from. The former factor can simply be addressed by investigating the dependence of spin dynamics on the number of Fe atom in the cluster. Starting from the same initial magnetic order, we performed spin dynamics calculations varying only the sizes of the clusters and keeping the other parameters constant. Using pulse amplitude of 5 T we have revealed a party effect in the magnetization reversal time, that is the switching time is smaller for odd number of Fe atoms whereas even number of Fe atoms show faster dynamics, see Fig. 4. This interesting phenomenon is related to the amount of energy accumulated while the magnetic pulse was on and consequently the relaxation of the magnetic energy. In the case of even number of Fe atoms the energy stored in the anisotropy is higher that the atomic spins are relatively stable and the relaxation process is faster.

It is worth to mention that selection of spin switching mechanism for given magnetic structure is complex process and certainly depends on multiple magnetic parameters. For instance, spin switching using only exchange forces, i.e. approaching the magnetically frustrated central atom with magnetic tip, it was impossible to switch orientation of the periphery atom in hexagonally ordered Fe atoms. Such experiment has performed for various ex- change distance between the tip and the hexagonal plane. The distance between the Fe atoms is so large that the lateral exchange force from the central atom is so weak that it could not magnetic reversal in all Fe atoms.

As summary we are able to reveal the multistable magnetic states in hexagonal arrays of Fe by tuning the shape of a magnetic pulse and able to determine the switching time. Our study reveals the optimum means of switching such magnetic superlattices by short pulses. We also analyzed the relaxation of the energy components (Zeeman, exchange and anisotropy) in time and explain the switching process thereof.